\documentclass[12pt,a4paper]{article}

\usepackage[T1]{fontenc}
\usepackage[utf8]{inputenc}
\usepackage[margin=2.5cm]{geometry}
\usepackage{amsmath,amssymb,amsfonts}
\usepackage{tabularx}
\usepackage{longtable}
\usepackage{array}
\usepackage{booktabs}
\usepackage{caption}

\usepackage{setspace}
\usepackage{authblk}
\usepackage[hidelinks,
            colorlinks=true,
            linkcolor=blue,
            citecolor=blue,
            urlcolor=blue]{hyperref}
\usepackage[numbers,sort&compress]{natbib}
\usepackage{url}

\title{\textbf{Beyond AI Delegation: A Prompt Pattern Framework for Productive Struggle and Evaluative Judgement in Secure Coding Education}}

\author[1]{Philipp Haindl}
\author[1]{Oliver Eigner}
\author[1]{Peter Kieseberg}

\affil[1]{%
  University of Applied Sciences St.\ P\"{o}lten,
  Campus Platz 1, 3100 St.\ P\"{o}lten, Austria}

\date{%
  \small
  Correspondence: \href{mailto:philipp.haindl@ustp.at}{philipp.haindl@ustp.at}\\[4pt]
  \normalsize
  \today}

\begin{document}

\maketitle

\begin{abstract}
\noindent\textbf{Purpose:}\enspace
Large language models make it easy for students to delegate writing,
analysis, and problem-solving to automated systems---bypassing the
effortful engagement that produces lasting understanding. We introduce
a practical framework that helps educators keep GenAI in the course
without removing the cognitive demands that make it worthwhile.

\medskip
\noindent\textbf{Methods:}\enspace
We apply Design Science Research (DSR) to synthesise and adapt a
taxonomy of nine prompt engineering patterns from established catalogs
in the computer science literature, mapped to two pedagogical
constructs---Productive Struggle and Evaluative Judgement. A course
design for an Advanced Secure Coding module, structured using the
DELTA framework, demonstrates the artifact's applicability.

\medskip
\noindent\textbf{Results:}\enspace
Nine prompt patterns, each mapped to a specific pedagogical function,
give instructors fine-grained control over how students interact with
AI\@. The secure coding design shows how three patterns---Flipped
Interaction, Alternative Approaches, and Cognitive Verifier---scaffold
vulnerability discovery and remediation while keeping students in the
reasoning role.

\medskip
\noindent\textbf{Conclusion:}\enspace
The framework provides a replicable approach to designing AI-augmented
learning experiences that preserve student reasoning, and establishes
a structured basis for future empirical evaluation in live course
settings.

\bigskip
\noindent\textbf{Keywords:} generative AI in higher education;
prompt engineering; productive struggle; evaluative judgement;
secure coding education
\end{abstract}

\newpage
\tableofcontents
\newpage

\section{Introduction}
\label{sec:intro}

Generative AI tools such as OpenAI ChatGPT, Anthropic Claude and Google
Gemini have arrived in university classrooms faster than most curricula
could adapt. The appeal is obvious: a well-posed question yields a
plausible essay, a working code snippet, or a structured analysis within
seconds---and students have noticed. One undergraduate put it bluntly in
a widely discussed Chronicle piece: using ChatGPT removes the need to
think~\cite{Terry2023}. The result is a delegation dynamic: cognitive
work that used to belong to the student is offloaded to the system, and
the student receives the output without having performed the reasoning
that produced it. When an AI can produce a complete solution, the student
no longer has to wrestle with the problem---and that struggle,
uncomfortable as it is, is where much of the learning
happens~\cite{Hiebert2007}.

This dynamic has a name in the education literature: \textbf{Productive
Struggle (PS)}. Hiebert and Grouws~\cite{Hiebert2007} describe it as
the phase in which learners grapple with tasks that exceed their current
understanding, forcing a reorganisation of knowledge structures that
would otherwise not occur. GenAI makes it easy to skip this phase
entirely. The risk is not that students use AI, but that they use it in
a way that produces outputs without producing understanding---a
distinction that matters a great deal in fields like computer security,
where shallow knowledge can have serious
consequences~\cite{Bozkurt2023,Southworth2023Model}.

The usual institutional response has been restriction: banning AI tools,
redesigning assessments to catch AI-generated text, or ignoring the
issue altogether~\cite{Chan2023}. None of these approaches is
sustainable. A more useful question is how to design courses in which AI
tools are present but do not replace the student's cognitive work.

This paper addresses that question. We argue that Prompt
Engineering---structuring the inputs to an LLM to obtain particular
kinds of output---is the missing link between pedagogical intent and
practical course design. Used naively, a prompt gets an answer. Used
deliberately, a prompt can force the student to analyze before the AI
explains, to compare before the AI recommends, to justify before the AI
confirms. The difference lies in how the prompt is constructed.

Our contribution is a taxonomy of nine prompt patterns, adapted from
software engineering catalogs~\cite{White2023,Pai2025}, and mapped to
two pedagogical goals: sustaining Productive Struggle and developing
\textbf{Evaluative Judgement (EJ)}---the capacity to assess the quality
of work, including AI-generated work, against appropriate
criteria~\cite{Bearman17082024}. We demonstrate how these patterns can
be applied in a master's-level secure coding module using the DELTA
course design framework~\cite{CTL-DeLTA}.

\section{Background}
\label{sec:rw}

\subsection{Productive Struggle and Evaluative Judgement}

The idea that learners benefit from working through difficulty is not
new. Vygotsky's zone of proximal development~\cite{Vygotsky1978} and
later empirical work by Hiebert and Grouws~\cite{Hiebert2007} both make
the case that effortful engagement with tasks just beyond current
competence is a prerequisite for deep, transferable understanding. The
practical implication is that removing the difficulty---whether through
tutors, textbooks, or now AI tools---removes the mechanism through which
understanding is built~\cite{Terry2023,Quach2022}.

Alongside Productive Struggle, Bearman et al.~\cite{Bearman17082024}
have argued for the growing importance of Evaluative Judgement: the
ability to assess the quality of one's own work and that of others,
including automated systems. In a context where AI can generate
plausible-sounding but incorrect or insecure code, the capacity to
evaluate critically is not a soft skill---it is a technical requirement.
Tai et al.~\cite{Tai2018} note that EJ develops through deliberate
practice of assessment and comparison, not passive consumption.
Lodge et al.~\cite{Lodge2023} make the related point that AI interactions
should be designed to augment, not replace, this kind of active
engagement.

\subsection{Prompt Patterns as a Structured Discipline}

Prompt engineering has matured considerably since the earliest
trial-and-error interactions with language models. White et
al.~\cite{White2023} and Schmidt et al.~\cite{Schmidt2024} were among
the first to apply the design patterns idea from software
engineering~\cite{Gamma1995} to the problem of LLM interaction. Rather
than crafting each prompt from scratch, a catalog of reusable
patterns---Persona, Template, Flipped Interaction, and so on---provides
a vocabulary of reliable techniques. Pai~\cite{Pai2025} later extended
this work with a six-category taxonomy that clarifies the intent behind
each pattern type.

These catalogs were developed primarily for software development
contexts. Our work adapts them for a different purpose. Schuelke-Leech
coins the term ``prompt literacy'' for the ability to craft effective
prompts, and argues that this is now a teachable component of AI
literacy~\cite{Jia2025}. We take that a step further: prompt literacy
is not just something students should learn \emph{about}---it is
something educators should use \emph{with} students to keep them in the
reasoning role when an AI is present.

\section{Methodology}
\label{sec:method}

The framework was deployed by the first author in a master's-level
research methods course at our university. We use \textbf{Design Science Research (DSR)} as the
methodological framework~\cite{hevner_design_2004}. DSR is a natural
fit for this work because our primary contribution is an artifact---the
prompt pattern taxonomy and the associated mapping to pedagogical
goals---rather than an empirical finding about student behaviour. The
goal is to produce something that practitioners can use and that future
research can evaluate and refine. Following Hevner et
al.~\cite{hevner_design_2004}, the work proceeded through four phases.

\subsection{Problem Identification}

We identified two related problems in contemporary HE: the erosion of
Productive Struggle as students offload cognitive work to GenAI, and
the absence of practical guidance that tells educators \emph{how} to
integrate AI without simply prohibiting it.

\subsection{Objectives}

The artifact should be specific enough to translate directly into course
activities, general enough to adapt across CS topics, and grounded in
documented pedagogical constructs rather than intuition.

\subsection{Artifact Design}

We worked in three stages. First, a review of the HE literature
identified PS and EJ as the competencies most directly threatened by
unrestricted GenAI use. Second, we compared the prompt pattern catalogs
of White et al.~\cite{White2023} and Pai~\cite{Pai2025}, selected the
patterns most relevant to educational settings, and adopted Pai's
six-category classification as the organising structure. Third, we
mapped each pattern to one or more pedagogical functions: scaffolding
analysis, eliciting comparison, compelling justification, or revealing
AI limitations---drawing an explicit line from each pattern to a
pedagogical purpose.

\subsection{Demonstration}

The framework was deployed in a master's-level research methods course.
Students learned the taxonomy explicitly and then used it to develop
research proposals for their theses, structured around the Design
Science Canvas~\cite{Johannesson2022}. This application gave us an
initial read on which patterns students found natural to use and which
required more scaffolding---informing the course design sketch in
Section~\ref{sec:course}.

\section{The Prompt Engineering Framework}
\label{sec:framework}

Our taxonomy is intended to sit \emph{inside} an existing course design
framework, not to replace one. Table~\ref{tab:frameworks} briefly
characterizes four frameworks that are relevant here. The DELTA
Framework~\cite{CTL-DeLTA} gets the most detailed treatment in
Section~\ref{sec:course} because it provides a natural home for the
prompt patterns at the activity design level.

\begin{table}[ht]
\caption{Pedagogical Frameworks Referenced in This Work}
\label{tab:frameworks}
\small
\begin{tabularx}{\textwidth}{@{}p{3.2cm}XX@{}}
\toprule
\textbf{Framework} & \textbf{Core Idea} & \textbf{Role in Our Approach}\\
\midrule
\textbf{Constructive Alignment}~\cite{Biggs1996}
  & Learning outcomes, activities, and assessment must be explicitly
    linked.
  & Ensures prompt-based activities directly address the competencies
    stated in the course outcomes.\\[6pt]
\textbf{3R-Framework}~\cite{StellenboschCTL3R}
  & Rethink, Redesign, Reassess: a cycle for adapting teaching practice.
  & Provides a high-level structure for iterating on AI-augmented
    course designs over successive offerings.\\[6pt]
\textbf{TPACK}~\cite{Mishra2006}
  & Effective teaching requires integrating Technology, Pedagogy, and
    Content knowledge.
  & Prompt engineering, in our view, belongs to the Technological
    Pedagogical Knowledge (TPK) layer.\\[6pt]
\textbf{DELTA Framework}~\cite{CTL-DeLTA}
  & A five-step iterative cycle: Context, Outcomes, Activities,
    Assessment, Reflection.
  & Used to structure the secure coding module in
    Section~\ref{sec:course}.\\
\bottomrule
\end{tabularx}
\end{table}

\subsection{A Taxonomy of Prompt Patterns}

Table~\ref{tab:patterns} lists the nine patterns in the taxonomy,
grouped into four categories adapted from Pai~\cite{Pai2025} and White
et al.~\cite{White2023}. For each pattern, the table gives the
educational rationale and a concrete example from a computing or
security context. The selection is not exhaustive; the original catalogs
each contain considerably more patterns. These nine were chosen because
they map directly onto activities where student cognition would otherwise
be most easily bypassed.

\begin{longtable}{@{}>{\bfseries\small\raggedright\arraybackslash}p{3.9cm}
                     >{\small}p{5.6cm}
                     >{\small}p{5.6cm}@{}}
\caption{A Taxonomy of Prompt Patterns for Pedagogical Application
  (adapted from White et al.~\cite{White2023} and
  Pai~\cite{Pai2025})}
\label{tab:patterns}\\
\toprule
\textbf{Pattern} & \textbf{Educational Rationale} &
\textbf{Example in CS\,/\,Security}\\
\midrule
\endfirsthead
\multicolumn{3}{c}{\small\tablename~\thetable{} -- \textit{continued}}\\
\toprule
\textbf{Pattern} & \textbf{Educational Rationale} &
\textbf{Example in CS\,/\,Security}\\
\midrule
\endhead
\midrule\multicolumn{3}{r}{\small\textit{Continued on next page}}\\
\endfoot
\bottomrule
\endlastfoot

\multicolumn{3}{@{}l}{\small\textit{Output Customization}}\\
\midrule
Persona
  & Setting a role shapes the AI's tone and stance, making it a
    consistent interlocutor rather than an answer
    machine~\cite{Pai2025}.
  & \textit{``Act as a skeptical code reviewer. Challenge my design
    decisions rather than confirming them.''}~\cite{mollick2023}\\[6pt]
Template
  & A fixed output format forces the student to parse and evaluate
    structured information~\cite{Pai2025}.
  & Requiring the AI to produce a vulnerability report in a
    standardised format (CWE ID, affected component, severity).\\[6pt]

\midrule
\multicolumn{3}{@{}l}{\small\textit{Error Identification}}\\
\midrule
Fact Check List
  & Asking the AI to list the factual claims underlying its answer
    makes those claims visible and checkable~\cite{Pai2025}.
  & \textit{``List every assumption you are making about the
    authentication flow before explaining the fix.''}\\[6pt]
Reflection
  & Prompting the AI to articulate its own reasoning exposes the gaps
    students should then
    investigate~\cite{Pai2025,Schmidt2024}.
  & \textit{``What are the limits of your analysis? What would you
    need to see to change your recommendation?''}\\[6pt]

\midrule
\multicolumn{3}{@{}l}{\small\textit{Prompt Improvement}}\\
\midrule
Cognitive Verifier
  & Breaking a problem into explicit steps (chain-of-thought) creates
    a visible reasoning trace the student can follow or
    challenge~\cite{Pai2025}.
  & \textit{``Walk through the refactoring step by step, explaining
    the security rationale at each stage.''}\\[6pt]
Question Refinement
  & The AI improves the student's own question, teaching problem
    formulation as a skill~\cite{Pai2025,Schmidt2024}.
  & A student submits a vague question; the AI returns a sharper
    version, which the student approves before proceeding.\\[6pt]
Alternative Approaches
  & Generating multiple options forces comparison and prevents
    students from accepting the first solution~\cite{Pai2025}.
  & \textit{``Propose three mitigation strategies and compare their
    trade-offs in a table.''}\\[6pt]

\midrule
\multicolumn{3}{@{}l}{\small\textit{Interaction}}\\
\midrule
Flipped Interaction
  & The AI asks questions rather than providing answers, keeping the
    student in the reasoning role~\cite{Pai2025,Schmidt2024}.
  & The AI conducts a Socratic interview, asking one question at a
    time until the student identifies a vulnerability
    independently.\\[6pt]
Game Play
  & A simulation frames the interaction as a scenario, sustaining
    engagement and introducing
    consequentiality~\cite{Pai2025,Schmidt2024}.
  & An incident-response simulation: the AI plays the attacker, the
    student plays the defender, decisions have modelled
    consequences.\\
\end{longtable}

\section{Applying the Framework: A Secure Coding Module}
\label{sec:course}

To show the taxonomy in action, we work through a complete course design
for a master's-level ``Advanced Secure Coding'' module, applying each
step of the DELTA Framework~\cite{CTL-DeLTA} in turn.

\subsection{Context}

The students in this course have solid programming backgrounds---they
can write functional code in Python or Java without assistance. What
they lack is experience reasoning about their code under adversarial
conditions. In current industry practice, they will write code alongside
AI assistants; research shows that such assistants frequently suggest
insecure implementations~\cite{Asare2023,Pearce2025}. The course must
therefore move beyond syntax and toward \emph{security-oriented code
review}: the capacity to recognise, analyse, and fix vulnerabilities
regardless of whether the code was written by a human or an AI.

\subsection{Learning Outcomes}

By the end of the module, students should be able to: (1)~identify
common vulnerability classes (SQL injection, XSS, insecure
deserialization) in code they did not write; (2)~compare remediation
strategies across at least two technical approaches and justify a
choice; (3)~write a security-oriented pull request description that a
colleague or auditor could act on; and (4)~recognise when AI-generated
code suggestions introduce new risk rather than resolving the original
one. The formulation matters: \emph{identify, compare, justify,
recognise} all require the student to actively reason, not merely to
generate.

\subsection{Learning Activities}

The central project runs over three activities, each scaffolded by a
different prompt pattern. Students work with a realistic web application
that contains a deliberately planted vulnerability.

\subsubsection{Guided Vulnerability Discovery}

Students must identify an SQL injection flaw in a Python Flask login
handler. Rather than asking the AI where the bug is, they use the
\textbf{Flipped Interaction} pattern:

\begin{quote}
\textit{``I am reviewing a Python Flask login function for security
vulnerabilities. Ask me questions one at a time to guide my analysis
until I can identify the specific type and location of the flaw.''}
\end{quote}

The AI asks about input sanitization, query construction, and error
handling. The student answers from the code. The flaw emerges through
the student's own analysis, not from an AI explanation---which means the
student cannot receive the answer without first demonstrating engagement
with the problem. Productive Struggle is preserved by design.

\subsubsection{Comparative Remediation}

Once the vulnerability is identified, students must choose a fix. They
use the \textbf{Alternative Approaches} pattern:

\begin{quote}
\textit{``Propose three distinct strategies to mitigate this SQL
injection vulnerability in Python. Compare the pros and cons of
(a)~parameterized queries, (b)~an ORM such as SQLAlchemy, and
(c)~a custom input sanitization function. Format the comparison as
a table.''}
\end{quote}

The student does not accept the AI's recommendation. Instead, they write
a justification for their chosen approach---this is where Evaluative
Judgement is explicitly exercised. The Template pattern constrains the
output format, making the comparison easier to evaluate.

\subsubsection{Supervised Implementation}

Students refactor the code using their chosen approach. The
\textbf{Cognitive Verifier} pattern keeps the AI from doing the work
silently:

\begin{quote}
\textit{``Break down the refactoring step by step. For each step,
state what you are changing and why it is more secure than the
original.''}
\end{quote}

The visible reasoning trace gives the student something to dispute or
confirm. It also provides a natural audit trail for assessment.

\subsection{Assessment}

The portfolio submission consists of four elements. First, the Flipped
Interaction chat log from Activity~1---the assessor can see whether the
student was guiding the analysis or simply agreeing with the AI.
Second, the written justification from Activity~2, which is graded on
the quality of the comparison and the coherence of the reasoning, not on
which approach was chosen. Third, the refactored code itself, checked
against the OWASP Top~10 for the relevant vulnerability class. Fourth, a
pull request description in a standard format (generated with Template
assistance), covering what was changed, why, and what testing would be
needed.

This design is deliberately resistant to academic dishonesty. A student
who outsources the entire project to an AI will produce a chat log that
shows no analytical process, a justification without genuine trade-off
reasoning, and code that may fix the original flaw but introduce others.
Each component is assessing something the AI cannot meaningfully fake on
the student's behalf.

\subsection{Reflection and Next Iteration}

After the first cohort, the instructor would review: Did students find
the Flipped Interaction appropriately challenging, or did they circumvent
it by restarting the conversation? Was the comparative table from
Activity~2 sufficiently nuanced, or did it become a formality? These
questions feed directly into the next design cycle---which is exactly
what the DELTA framework asks for.

\section{Limitations}
\label{sec:limitations}

Following established practice in design science
evaluation~\cite{hevner_design_2004}, we structure our limitations
around standard threats to validity, applied here to the artifact
demonstration rather than a controlled experiment.

\paragraph{Construct Validity.}
The capabilities of commercial LLMs change
quickly~\cite{Southworth2023Model}. A pattern like Flipped Interaction
works precisely because current models respond well to instruction-based
constraints. If future models infer pedagogical context and apply these
constraints automatically, the explicit patterns may become
unnecessary---or new ones will be needed. The taxonomy should be treated
as a versioned artifact, not a fixed catalog. Whether the constructs of
Productive Struggle and Evaluative Judgement are faithfully
operationalized by the patterns as described also remains to be verified
through empirical observation of student behavior.

\paragraph{Internal Validity.}
The secure coding module described in Section~\ref{sec:course} is a
design artifact. The DSR demonstration phase used a different course
context---a research methods module where students applied the taxonomy
to thesis proposal development---and did not include controlled
measurement. We therefore cannot report on student performance,
engagement, or the actual difficulty of the activities as experienced by
learners, nor can we rule out alternative explanations for any observed
effects in a future deployment.

\paragraph{External Validity.}
The demonstration deployment was conducted in one master's programme at
one university. We cannot speak to how the framework would transfer to
undergraduate courses, other institutional contexts, or disciplines
outside computing. The prompt patterns themselves are not CS-specific,
but whether the same mapping to pedagogical goals holds elsewhere remains
to be tested.

The next step is a controlled deployment of the secure coding module,
with pre- and post-measurement of security reasoning ability, collection
of the chat logs as a data source, and structured student feedback on
the pattern-based activities. That evaluation is planned for the
following academic year.

\section{Conclusion}
\label{sec:conclusion}

Whether GenAI helps or hinders student learning depends less on the tool
itself and more on how the prompts are designed. Given a poorly chosen
prompt, an AI removes the need for the student to think. Given a
well-chosen one, it can force exactly the kind of thinking the course is
trying to develop.

The taxonomy we have presented gives instructors a vocabulary for making
that choice deliberately. Nine patterns, four categories, each one
mapped to a specific pedagogical function. The secure coding module shows
what it looks like in practice: three activities, three patterns, each
one preserving a different aspect of the reasoning process that
assessment would otherwise be unable to see.

The framework is a first iteration. The conceptual work---mapping
engineering patterns to pedagogical goals---is, we believe, a useful
contribution in its own right, and one that has been missing from the
debate about AI in computer science education.


\section*{Declarations}

\paragraph{Competing Interests.}
The authors declare no competing interests.

\paragraph{Funding.}
No funding was received for this study.

\paragraph{Ethics Approval.}
The demonstration described in this paper was conducted as part of
regular teaching activity at University of Applied
Sciences St.\ P\"{o}lten. No personal data were collected, no experimental intervention
beyond normal course delivery was applied, and no individual results are
reported. Formal ethics approval was therefore not required.

\paragraph{Data Availability.}
No datasets were generated or analysed during the current study.

\paragraph{Author Contributions.}
Conceptualization: P.H., O.E., P.K.;
Methodology: P.H.;
Investigation: P.H.;
Writing -- original draft: P.H.;
Writing -- review \& editing: P.H., O.E., P.K.


\bibliographystyle{unsrtnat}
\bibliography{references}

\end{document}